# The use of strain and grain boundaries to tailor phonon transport properties: A first principles study of 2H-phase CuAlO$_2$ (Part II)

Evan Witkoske[1], Zhen Tong[4], Yining Feng[3], Xuilin Ruan[4,5], Mark Lundstrom[1], and Na Lu[2,3,5]

[1]School of Electrical and Computer Engineering, Purdue University, West Lafayette IN, 47907
[2]School of Materials Engineering, Purdue University, West Lafayette, IN, 47907
[3]Lyles School of Civil Engineering, Purdue University, West Lafayette, IN, 47907
[4]School of Mechanical Engineering, Purdue University, West Lafayette, IN 47907
[5]Birck Nanotechnology Center, Purdue University, West Lafayette, IN 47907

**Abstract:** Transparent oxide materials, such as CuAlO$_2$, a p-type transparent conducting oxide (TCO), have recently been studied for high temperature thermoelectric power generators and coolers for waste heat. TCO materials are generally low cost and non-toxic. The potential to engineer them through strain and nano-structuring are two promising avenues toward continuously tuning the electronic and thermal properties to achieve high zT values and low $cost/kW-hr devices. In this work, the strain-dependent lattice thermal conductivity of 2H CuAlO$_2$ is computed by solving the phonon Boltzmann transport equation with interatomic force constants extracted from first-principles calculations. While the average bulk thermal conductivity is around 32 W/(K-m) at room temperature, it drops to between 5-15 W/(K-m) for typical experimental grain sizes from 3nm to 30nm at room temperature. We find that strain can offer both an increase as well as a decrease in the thermal conductivity as expected, however the overall inclusion of small grain sizes dictates the potential for low thermal conductivity in this material.

## I. Introduction

Thermoelectric (TE) devices and materials are appealing for use in solid-state energy generation and solid-state cooling. Regardless of their operating mode, a good thermoelectric material should have a high electrical conductivity ($\sigma$), Seebeck coefficient (S), and a low lattice thermal conductivity ($\kappa_L$) given in the figure of merit [1]

$$zT = \frac{\sigma S^2 T}{\kappa_L + \kappa_e}. \qquad (1)$$

However robust and reliable as TE devices could potentially be, they have been limited by low conversion efficiencies since the beginning [2]–[6]. The gains in *zT* have been primarily driven by a reduction in the lattice thermal conductivity of materials and devices through the use of nano-structuring [7]–[13] and the development of novel thermoelectric materials with the ability to take advantage of a wide range of operating temperatures [14]–[17] with inherently low thermal conductivity. These advances have not translated into working devices [18] however, due to many issues, one of which being material and fabrication cost. As we approach the lower limit of the lattice thermal conductivity for complex TE materials the applicability of the field of thermoelectrics remains in question due to the cost and efficiency of working devices.

A previous work [19] looked at transparent conducting oxides (TCOs), and specifically 2H-phase $CuAlO_2$, which has gained interest as a promising candidate for high temperature p-type thermoelectric applications [20]–[23] [24], because of its potential use in high temperature applications, due to a large band gap, high thermal stability, oxidation resistance, and low material costs [20], [21], [25]–[34]. Some experimental and theoretical studies [19], [20], [24], [35]–[37], have been done on the thermoelectric (TE) properties of the 2H phase of this material, however none have looked at the thermal conductivity using rigorous first principles simulations. Under specific cases of strain, n-type conduction can produce higher power factors than their p-type counterparts providing an interesting avenue for strain engineering to produce both n and p type legs from the same material [19]. Strain may be beneficial for the electronic component of $zT$, but its effect on the lattice thermal conductivity must also be ascertained. That is the objective of this work.

In practice, defects are introduced into the lattice itself as point defects or as grains in micro-structured thermoelectric materials to aid in suppressing thermal conductivities. The creation of grain boundaries through the use of nano-structuring is one of the most promising and widely used strategies to improve zT [38]. In this work, a variety of nanometer grains sizes will be simulated, that are consistent with the small nano-scale size limit (SNS) [39], [40], giving rise to grain boundary scattering [41] of nano-structured TE materials. We show in this work an ab initio assessment of lattice thermal conductivity in 2H $CuAlO_2$, including third-order anharmonic scattering, natural isotopic scattering, and Casimir finite-sized boundary scattering, which takes into account the spectral decomposition of phonon wave vectors at the grain boundary. We find that the low thermal conductivity seen experimentally [42], [43] is most likely due to micro and nano-structured effects due to grain boundary scattering of phonons. Casimir grain boundary scattering reduces the thermal conductivity by as much an order of magnitude, with crystalline anisotropy (due to the hexagonal structure) further reducing $\kappa_L$. Isotopic scattering has a limited effect (especially for small grain sizes) due to the constituent atoms being on the smaller end of the periodic table, (typically isotopic scattering plays a larger role when large mass atoms are involved or at low temperatures [44]).

In this work under a variety of strains are applied and their effects on the lattice thermal conductivity will be discussed, as well as the variations of the lattice thermal conductivity for a wide range of temperatures. There are five sections in this paper, I) introduction, II) atomic structure and methodology, III) unstrained lattice thermal conductivity, IV) strained lattice thermal conductivity and finally, V) conclusions.

## II. Methodology and Simulation details

### A. Atomic structure

$CuAlO_2$ crystallizes in two distinct phases, 3R and 2H, both having a delafossite structure with the rhombohedral (3R) and hexagonal (2H) phases occurring at atmospheric pressures [45]. In Fig. 1(a), the 2H phase structure, with a space group of $P6_3/mmc$ (no. 194), is shown with the crystallographic directions "x, y, and z". Figure 1(b) shows the high symmetry k-points of the first Brillouin zone, which are used for plotting the phonon band structures.

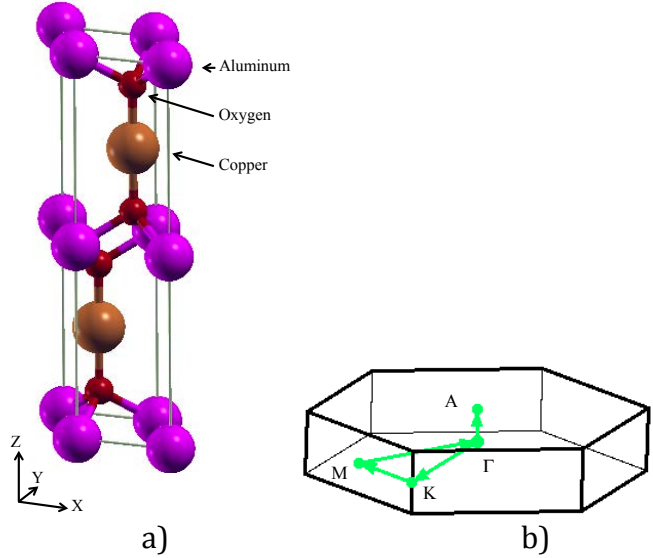

Figure 1. (a) Relaxed conventional supercell of 2H-phase $CuAlO_2$. (b) First Brillouin zone with the high symmetry points used for the dispersion paths shown in Fig. 2.

A phonon can be defined as a quantum of lattice vibrations, described by the quantum number $\lambda = (v, \mathbf{q})$, where $v$ denotes the branch index and $\mathbf{q}$ denotes the wave vector of a particular phonon mode. A phonon can be scattered through interaction with other phonons, electrons, impurities, grains, etc. The overall scattering rate of a phonon mode can be estimated by Matthiessen's rule [46] as $\Gamma_\lambda = \Gamma_\lambda^{pp} + \Gamma_\lambda^{pe} + \Gamma_\lambda^{gb} + \Gamma_\lambda^{im}...$ that includes phonon-phonon (p-p) scattering, phonon-electron (p-e) scattering, grain boundary (gb) scattering, phonon-impurity (isotopic) scattering, respectively. Due to the low electrical conductivity as noted in [19], [25], [42], the phonon-electron scattering term, $\Gamma_\lambda^{pe}$, is neglected here. Isotopic scattering was included in some simulations, however the results show that the effect is negligible, especially when grain sizes are small. Only $\Gamma_\lambda^{pp}$, and $\Gamma_\lambda^{gb}$ will be calculated from first principles in this work.

The current calculation includes three-phonon scattering only, while four-phonon scattering was shown to be important for certain materials [47], [48]. Further studies of this material warrant the inclusion of four-phonon processes since present applications of $CuAlO_2$ are for high temperatures, where four-phonon scattering can be relevant [49]. The p-p scattering contribution from three-phonon processes to $\Gamma_\lambda^{pp}$ is given by Fermi's golden rule as [50], [51],

$$\Gamma_\lambda^{pp} = \frac{\hbar\pi}{4N} \sum_{\lambda_1\lambda_2}^{+} \frac{2(n_1 - n_2)}{\omega\omega_1\omega_2} \left|V_{\lambda\lambda_1\lambda_2}^{+}\right|^2 \delta(\omega + \omega_1 - \omega_2)$$
$$+ \frac{\hbar\pi}{8N} \sum_{\lambda_1\lambda_2}^{+} \frac{(n_1 + n_2 + 1)}{\omega\omega_1\omega_2} \left|V_{\lambda\lambda_1\lambda_2}^{-}\right|^2 \delta(\omega - \omega_1 - \omega_2), \quad (2)$$

where the first term is the combination of two phonons with possibly different wave vectors to produce one phonon. The second term is for the opposite process of one phonon splitting into two, i.e. phonon emission, with $\hbar$ being Planck's constant, $n_i$ is the Bose-Einstein distribution, and $\omega$ is the phonon frequency. The summation runs over all phonon modes and wave vectors and requires conservation of the quasi-momentum $q_2 = q \pm q_1 + Q$ in which Q is the reciprocal lattice vector with Q = 0 constituting a normal process and $Q \neq 0$ being an Umklapp processes. N is the number of discrete q-points of the $\Gamma$-centered q-grid for sampling, $\delta$ is the Dirac delta function, which is approximated by a Gaussian function in the computational package QUANTUM ESPRESSO (QE) [52], [53] (which is the computational package used in this work). The scattering matrix elements $V_{\lambda\lambda_1\lambda_2}^{\pm}$ [50], [53], [54] are given by,

$$V_{\lambda\lambda_1\lambda_2}^{\pm} = \sum_{l_1}^{N_B} \sum_{l_2,l_3}^{N,N} \sum_{\eta_1\eta_2\eta_3}^{3,3,3} \frac{\partial^3 E}{\partial r_{l_1}^{\eta_1} \partial r_{l_2}^{\eta_2} \partial r_{l_3}^{\eta_3}} \frac{e_\lambda^{\eta_1}(l_1) e_{j_1,\pm q_1}^{\eta_2}(l_2) e_{j_2,-q_2}^{\eta_3}(l_3)}{\sqrt{m_{l_1} m_{l_2} m_{l_3}}}, \quad (3)$$

where $m_{li}$ is the atomic mass and $e_{\nu,q}$ is a normalized eigenvector of the mode $\lambda = (\nu, \mathbf{q})$. In Eq. (3) $l_1$, $l_2$, and $l_3$ run over all atomic indices (with $l_3$ only summing over the atoms in the center unit cell, which contains $N_B$ basis atoms), and $\eta_1$, $\eta_2$, and $\eta_3$ represent Cartesian coordinates. The 3rd order anharmonic interatomic force constants (IFCs) are the third-order partial derivatives, which are obtained from the D3Q-Thermal2 package interfaced to QE [53], [54]. The energy $E$ is the total energy of the entire system with $r_{l_1}^{\eta_1}$ designating the $\eta_1$ component of the displacement of a particular atom $l_1$. We also obtain the second order force constants by Fourier transforming dynamical matrices in the reciprocal momentum space gleaned from linear response theory implemented in QE [53].

The lattice thermal conductivity tensor can then be calculated as

$$\kappa_{L,\alpha\beta} = \sum_\lambda \frac{1}{k_B T^2} (\hbar\omega_\lambda)^2 n_\lambda (n_\lambda + 1) \upsilon_{\lambda,\alpha} \upsilon_{\lambda,\beta} \tau_\lambda, \quad (4)$$

$$\frac{1}{\tau_\lambda} = \frac{1}{\tau_{\lambda,pp}} + \frac{1}{\tau_{\lambda,gb}}, \quad (5)$$

with Matthiessen's rule being applied to the scattering lifetimes of individual phonon modes separately [55].

**B. Grain boundaries**

The phase or polytype transition of this material from the trigonal 3R to the hexagonal 2H phase can happen at 15.4 Gpa [37], with the 3R phase being studied more due to its better structural stability. It has been shown elsewhere that treating grain boundaries as a secondary phase of a material can help explain much of the transport behavior observed in polycrystalline samples [56]. At grain/phase boundaries that are comparable to grain size, a significant amount of heat is transported through the interface by phonons [56], therefore studying the high pressure 2H phase of $CuAlO_2$ can help elucidate phonon transport in nano-grained structures of the 3R phase as well.

The grain boundary scattering term is the only term that depends on the direction of the phonon group velocity explicitly, which can generally be expressed as the frequency independent equation [57],

$$\tau_{\lambda,gb} = \frac{L_\eta}{2|v_{\lambda,\eta}|}, \qquad (6)$$

where $L_\eta$ is the distance between the two boundaries in one of three Cartesian directions. However due to the omnidirectional grains and the variety of temperature ranges used, a model from [58] is assumed in this study, which assumes a grain boundary acts as a diffraction grating, producing diffraction spectra of various orders. Multiple values of $L_\eta$ are used in this work and are shown in Section III. In all calculations, including Casimir scattering, we assume the correction factor [53] [59], [60] to be F =1, which takes into account the shape and roughness of each grain boundary to be diffusive.

It has been shown in previous work the commonly used frequency-independent boundary scattering (grey model or the simple Casimir model, not to be confused with the Casimir model used in this work [58]) can better fit thermal conductivity experimental data if it is replaced by a frequency-dependent phonon scattering model due to dislocation strain in grain boundaries [61]. This difference manifests itself most notably at low temperatures with the frequency-independent boundary scattering going as the normal $\sim T^3$ Debye law, which deviates from experimental thermal conductivity of polycrystalline silicon which goes as $\sim T^2$ at low temperatures [62]. In this work however, the temperature range of interest is primarily >300K, thus these discrepancies are not considered here, and will be left for future work on this material and its experimental low temperature thermal conductivity behavior.

## C. Simulation details

The thermal conductivity, q–dependent linewidths, including Casimir and natural isotopic-disorder scattering, are calculated using the D3Q-Thermal2 codes [53], [54]. We use a generalized-gradient approximation (GGA) Perdew-Burke-Ernzerhof (PBE) scheme [63] for electron-electron interaction, and ONCV norm-conserving pseudopotentials [64] for the electron-ion interaction. GGA considers the gradient of the charge density at each position when the atom position is perturbed, and has been shown to work better for materials with abrupt charge density changes such as in semiconductors, whereas the Local Density Approximation (LDA) is more applicable to metallic systems [65].

The plane-wave energy cut-offs and force thresholds for the variety of strain and unstrained cases were varied based on finding well-relaxed structures with the absence of negative phonon frequencies and are provided in section one of the supplementary material (SM). The hexagonal symmetry was enforced during the geometry optimization. Strained structures were relaxed after artificially changing the lattice constant in a particular direction of strain. The k–point grids used for structural relaxation and optimization were 16 x 16 x 16 while the q-point grids for phonon dynamical-matrix calculations were set to 4 x 4 x 4. For the third-order force constants a grid of 2 x 2 x 2 was used, and a 10 x 10 x 10 grid was used for the lattice thermal conductivity calculation. In calculating the lattice thermal conductivity, a Gaussian smearing of 5 cm$^{-1}$ was used. This value, along with tests on k/q-grids, and energy cut-offs, have been checked w.r.t the phonon frequencies at the Γ point and suggest that the average lattice thermal conductivity value is stable to within ten percent. Calculations on the convergence of these parameters can be found in section two of the SM.

## D. Anisotropy and Convergence

In this work, due to the "mis"-orientation of grains in experimental samples [42], the uncertainty in the heat and electronic directed transport intended, the bulk nature of the intended material, and the "relative" isotropy in the lattice thermal conductivity tensor, the lattice thermal conductivity is assumed to be $\kappa_L = \sum_{i=1}^{3} \kappa_i / 3$, i.e. an average over the x, y, and z directions. For the lattice thermal conductivity calculation, the "exact" iterative conjugate-gradient solution (CGS) method of [53] which attempts to solve the linearized BTE exactly, was compared to the single mode relaxation time approximation (SMRTA) [53], [66]–[69]. The SMRTA is considered to be inadequate to describe thermal conductivity at low temperature ranges, with increased sensitivity to isotopic scattering at these low temperatures[50], [70] i.e. T < 300 K. However for our purposes here, the temperature range of interest are generally greater than 300 K, even though results will be shown for temperatures less than this, the SMRTA gives results above 300 K for the average thermal conductivity "consistent" with the CGS. For structures with no grain scattering assumed, the lattice thermal conductivity for the x and y directions are consistent within 1-5% for the SMRTA compared to the CGS for all temperatures considered. However, the lattice thermal conductivity for the z direction shows a large deviation for T > 300 K for the SMRTA compared to the CGS, with the SMRTA model underestimating the z directed lattice thermal conductivity by around 40-50% for these temperatures. The results of the direction resolved thermal conductivity are shown in section two of the SM. Even with this large deviation, the average lattice thermal conductivity found by averaging the x, y, and z directions deviates only around 10% from the SMRTA to the CGS for T > 300 K. Due to the considerations mentioned at the beginning of this section, and, as we will see later, that deviations are significantly reduced when small grain boundaries are introduced into the structures, the average thermal conductivity using the SMRTA is used throughout this work.

### III. Unstrained Thermal conductivity

The lattice constants in the hexagonal relaxed structure were found to be a = b = 2.8798 Å and c = 11.4077 Å, which agree well with experimental [71] and theoretical [35], [71] results.

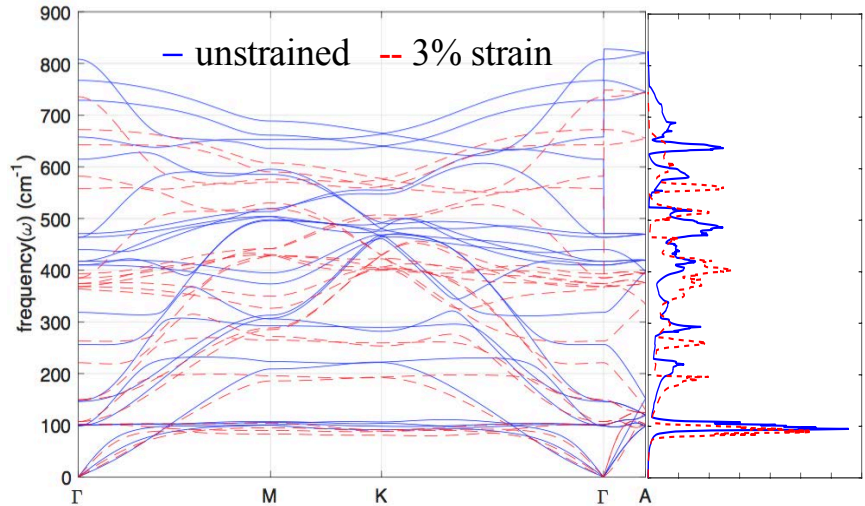

Figure 2. Calculated phonon dispersion relation and density of states of the 2H phase of $CuAlO_2$ for unstrained (solid blue) and +3% strain (dotted red).

Figure 2 shows the phonon band dispersion for 2H $CuAlO_2$ for the unstrained structure (and +3% strain for comparison) used throughout this section. With the hexagonal structure of this material, one notices the large amount of bands especially around the low frequency of 100 $cm^{-1}$. Boundary or grain scattering generally scatter low frequency phonons, as opposed to Umklapp and point defect mechanisms which scatter at all frequencies and high frequencies respectively [61]. We will see that these low frequency modes in 2H $CuAlO_2$ get effectively scattered when grain boundaries are introduced. A plot of the phonon linewidths superimposed on the band structure can be found in section two of the SM.

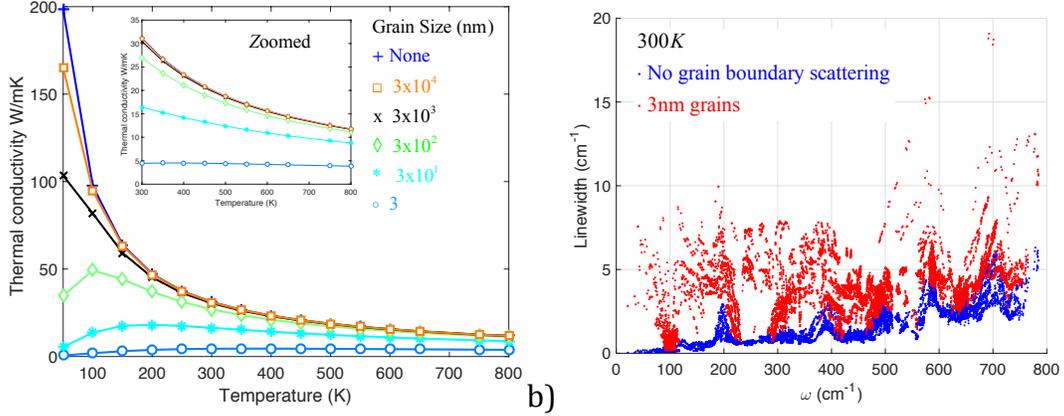

Figure 3. (a) Comparison between the thermal conductivity of the material without grains, along with grain sizes of 3nm, 30nm, 300nm, 3000nm, and 30000nm vs. temperature. (b) Scattering linewidth vs. frequency for the structure with no grains (blue dots) and the structure with grains of 3nm (red dots).

The structures in Fig. 3(a) show a $\sim T^{-1}$ behavior above the Debye temperature, consistent with the Dulong-Petit law, with the thermal conductivity being governed by the MFP in this region. With the finite size of the crystal being accounted for, $\kappa_L$ becomes finite at zero K and decreases for grains larger than 3000nm as temperature increases. However, for structures with grains of 300nm and less, the lattice thermal conductivity begins to increase from 0K before then decreasing around the Debye temperature, and continuing to do so at higher temperatures. As was mentioned earlier, the difference between the $\sim T^3$ Debye law for low temperature phonons due to boundary scattering with the dislocation strain model, $\sim T^2$, won't be resolved in this work.

This can be interpreted as a dominating effect of the grain boundaries over intrinsic phonon scattering [72] (i.e. the intrinsic mean free path, $l_{mfp} = \bar{v}\tau_\lambda$ is very large at low temperatures and is limited by the crystal size). The same (albeit difficult to see) effect can be observed in Fig. 3(b) at 300 K. Even though the linewidths have greater values for some of the higher frequency modes, the lower frequency linewidths for the 3nm grain case are consistently larger in comparison to the no grain structure (blue dots) at these low frequencies compared to the discrepancy between the two structures for higher frequencies. These low frequency bands are generally affected more by Casimir scattering, with both anharmonic and Casimir scattering being significant along the shearing transverse-acoustic mode around 100 cm$^{-1}$, as can be seen in Fig. 1 around the $\Gamma$ point.

This same effect can be seen in the accumulation functions [68] in Fig. 4(a) (no grains) and Fig. 4(b) (3nm grains) as well with lower temperatures being governed by larger MFPs for both the structure without grains Fig. 4(a) and for 3nm grains Fig. 4(b).

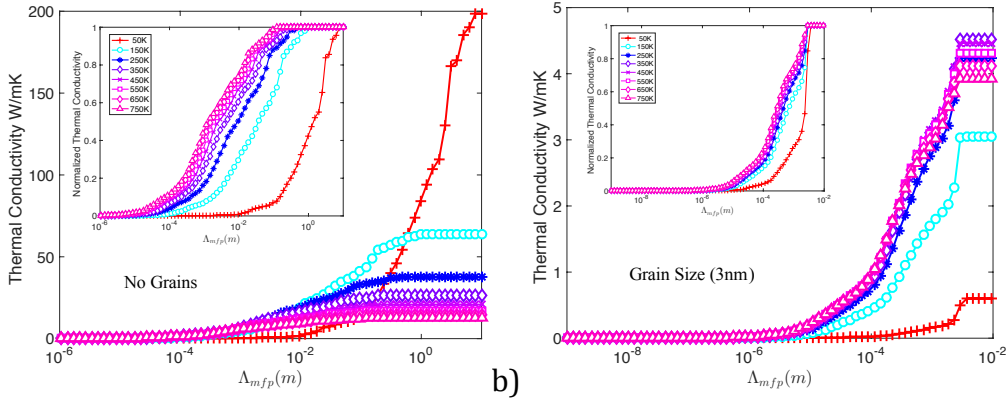

Figure 4. (a) The accumulation function of the thermal conductivity for a structure with no grains for a variety of temperatures. The inset is the thermal conductivity normalized to 1. (b) The same set of plots but for a structure that includes 3nm grains. Please note the different values on the x-axis.

In comparing with experiment at 300 K, values of the total thermal conductivity agree well with reported values for "micro" grained structures [43] of 20-30 W/mK, with our results 25-32 W/mK being reasonable. For "nano" structured samples in [43], around 2 W/mK was found compared to our results of 4.4 W/mK . If isotopic scattering is included with natural isotopes assumed, our thermal conductivity is only reduced to around 3.8 W/mK, making the isotopic scattering contribution negligible. Further results were obtained from [42] with thermal conductivities found to be around 2.5 W/mK at 300K. The grain size observed in [42] was experimentally found to be 0.247 nm. Though not included in the above figures, a grain size of 0.3nm in our simulations gives a thermal conductivity of around 0.5 W/mK at 300 K. The trend seen in Fig. 3(a) however for the 3nm structure is consistent with the thermal conductivity seen in [42], which is relatively constant over a large temperature range from 300K-800K.

## IV. Thermal Conductivity with Strain

### A. Hydrostatic strain

To simulate hydrostatic strain (equal strains in all directions), we took the relaxed structure and applied isotropic strain to the cell parameters by ±1, ±2, and ±3%. The atomic positions were then allowed to relax keeping the volume of the cell constant. Imparting confidence in this particular methodology, the lattice parameters under these hydrostatic strains are found to be consistent with theory and experimental values [73], [74].

Figure 5(a) is a plot of the thermal conductivity vs. strain for both a structure without grains (blue circles) and for one with 3 nm grains (red stars). We can see from Fig. 5(b) that the +3% strained structure without grains has a much higher scattering linewidth, with values being shifted lower in frequency, and conversely for the -3% strain, the linewidths are lower and shifted higher in frequency. We remind the reader that -3% constitutes compressive strain.

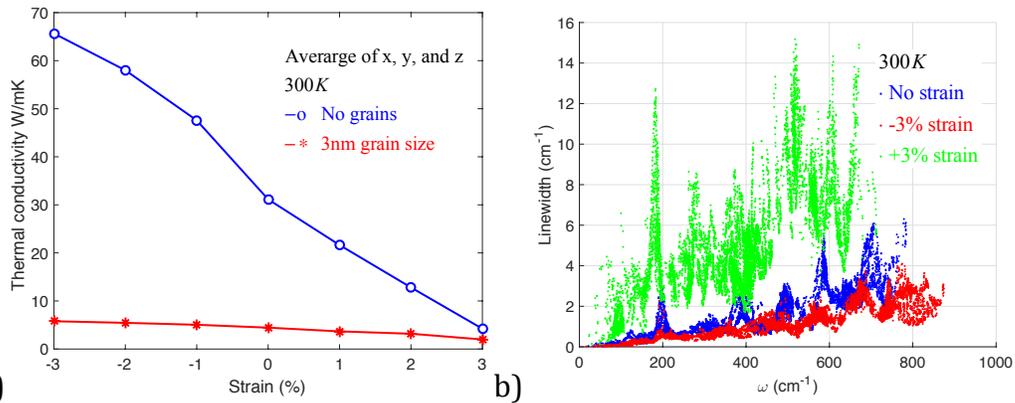

Figure 5. (a) Lattice thermal conductivity vs. strain for both a structure without grains (blue circles) and for one with 3 nm grains (red stars). (b) Linewidth vs. frequency for structures with no grains with the unstrained case as blue dots, -3% strain shown as red dots, and +3% shown with green dots.

Figure 5.5(b) explains Fig. 5.5(a), with the large linewidths corresponding to a shorter lifetime, yielding $\kappa_L$ for a structure with no grains with +3% strain comparable to that of a structure with 3nm grains at +3% strain (4 W/mK compared to 2 W/mK). The $\kappa_L$ at +3% strain without grains at 300K is about 4 W/mK compared to the unstrained case at 300 K of 32 W/mK. The reduction in lattice thermal conductivity in Fig. 5.5(a) from the -3% strained to the unstrained case of 55% is conversely consistent with the reduction of the average linewidth in Fig. 5.5(b) of ~50% from the unstrained to -3% case. The same effect is seen from the unstrained to +3% strain cases, with the lattice thermal conductivity reduced by 86% in Fig. 5.5(b), with the average linewidth being reduced by around ~70% from the +3% to unstrained case. Even though these trends aren't exactly one to one, the connection to scattering is clear. The trend vs. strain is consistent with compressive strain shifting the phonons to higher frequencies with bands becoming more spread out in energy inducing longer scattering times, giving higher thermal conductivities. The +3% strained phonon energy dispersion bands are lower than the unstrained case and closer together, as seen in Fig. 5.2, therefore, the lower frequency induces stronger phonon-phonon scattering (predominately among high frequency optical phonons) in the strained case, as seen in Fig 5.5(b). Structures with 3nm grains are also reduced by the inclusion of strain, varying from around 6 W/mK to 2 W/mK in a similar manner.

What is more interesting is the trend observed in Figs. 6(a) and 6(b). Figure 6(a) is a plot of $\kappa_L$ vs. strain for a wide range of temperatures. The thermal conductivity in Fig. 6(a) decreases from -3% to +3% just as before, as does $\kappa_L$ for the temperature trend, i.e. the lowest temperature has the highest $\kappa_L$ and progressively downward with values converging to less than 20 W/mK for all temperature cases. However Fig. 6(b), which is a plot of $\kappa_L$ vs. strain for a structure with 3nm grains, has the lowest thermal conductivity for the lowest temperature curve 50 K. However, the trend is not entirely flipped; the highest $\kappa_L$ is not the highest temperature at -3% strain, but a mid range temperature of around 450 K. Further more at +3% strain the curves begin to cross with the highest $\kappa_L$ around 2 W/mK for 300 K and the lowest still being 50 K at 0.5 W/mK, with the highest temperature of 800 K occupying the second lowest value of slightly over 1 W/mK.

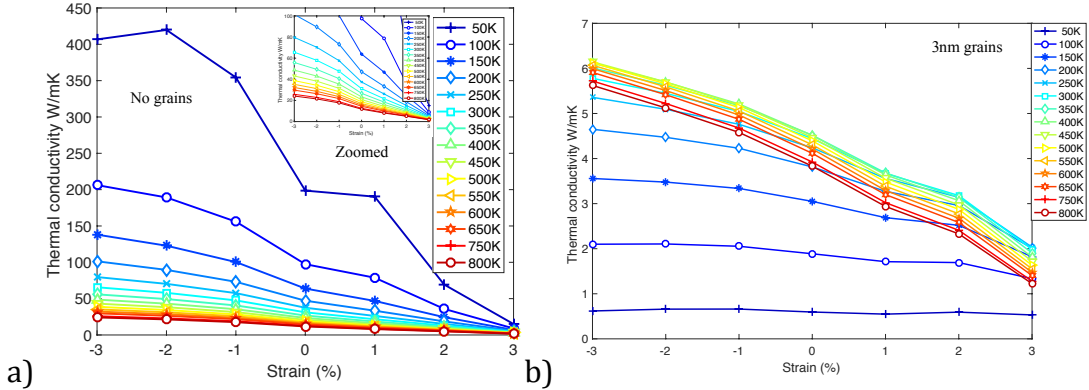

Figure 6. The figures above show plots of the thermal conductivity vs. hydrostatic tensile strain ranging from -3% to 3% with a variety of temperatures with symbols in the figures; (a) is without grains in the structure, and (b) has 3nm grains included. Both plots have the same symbols for each temperature case. The inset in (a) is zoomed to thermal conductivities from 0 to 100 W/mK.

As was mentioned before in the methods section, the SMRTA is generally less accurate for lower temperature ranges up to 300 K, with a general reduction in $\kappa_L$ for this method compared to the conjugate-gradient solution (CGS) method [53]. Neglecting the low temperatures, we see that for temperatures from around 300 K and above, the $\kappa_L$ values are grouped closer together in both figures, more noticeably in Fig. 6(b) ranging from around 5.5-6 W/mK for -3% strain, to around 2 W/mK for +3% strain. These trends suggest that nano-structured $CuAlO_2$ with small grain sizes of around a nanometer are temperature independent regardless of the hydrostatic strain applied, also seen in Fig. 3(a), further clarifying the constant temperature curve of the experimental data in [42] and [43].

### B. Selected uni-axial strain

In part I of this study [19], it was found that +1% strain in the z direction induced the highest power factors for both n and p type transport in $2H-CuAlO_2$. A structure with +1% strain in the z direction was simulated for both a crystalline (no grains) as well as for a structure with 3nm grains. The crystalline structure and polycrystalline 3nm grain structure had identical temperature trends to that observed in figure 3(a) for the same cases. As far as lattice thermal conductivity is concerned, for this material it is generally unnecessary to resolve uni- or bi-axial strain, since the 3nm grains will dominate the scattering and the results are the same as above. The large reduction of 30 – 40% in lattice thermal conductivity due to 1 – 3% hydrostatic strain is consistent with first principle work in other materials [75]. The inclusion of only uni-directional strain has a much smaller effect of 1 – 10% on the overall lattice thermal conductivity, and many times not at all.

**Conclusions**

In summary, we believe that the lattice thermal conductivity and therefore the total thermal conductivity of 2H CuAlO$_2$, due to the limited electronic thermal conductivity, is not unnaturally small because of its inherent structure, but rather because of the inclusion of grain-boundary scattering. While the crystalline thermal conductivity is around 32 W/(K-m) at room temperature, it drops to between 5-15 W/(K-m) for typical experimental grain sizes from 3nm to 30nm at room temperature. A second conclusion of this study shows that when grains of 3nm or less are assumed, the thermal conductivity is generally independent of temperature for unstrained, compressive, or expansive strained structures. Due to the array of possible strain and phase transitions inherent in grain boundaries, these results confirm similar experimental studies that show limited temperature dependence for $\kappa_L$ regardless of the fabrication technique used. In the particular case of +3% strain on a crystalline sample, the lattice thermal conductivity is similar to that of samples with 3nm grains. However, due to difficulties in fabrication and the omnidirectional nature of grains in most experimental samples of this material, it will be more practical to decrease the lattice thermal conductivity through the inclusion of these grains, than by straining crystalline samples.

In conjunction with previous studies of 2H CuAlO$_2$, there is a possibility for use of this low cost and non-toxic transparent conducting oxide (TCO) as a TE generator for both high and room temperature applications by fabricating polycrystalline structures with nano-scale features. Although the electrical conductivity of this material is rather low, and the overall *zT* efficiencies may not exceed state of the art TE materials, the $cost/kW-hr quite possibly could.

**Acknowledgements** – This work was partially supported by NSF CAREER project (CMMI 1560834)

***Data Availability*** – The pseudo-potentials used in the DFT calculations can be found at http://www.quantum-espresso.org/pseudopotentials.